\documentclass[fleqn,usenatbib,onecolumn]{mnras}

\usepackage[T1]{fontenc}
\usepackage{ae,aecompl}

\usepackage{xcolor}
\usepackage{geometry}
\usepackage{natbib}
\usepackage{pifont}
\usepackage{dcolumn}
\newcolumntype{d}[1]{D{.}{.}{#1}}

\usepackage{graphicx}	%
\usepackage{amsmath}	%
\usepackage{amssymb}	%
\usepackage{pdflscape} %
\usepackage[version=4]{mhchem}

\newcommand{\icm}{cm\textsuperscript{-1}}

\usepackage{newtxtext,newtxmath}

\title[Revised benzene binding energy]{Hybrid approach predicts a lower binding energy for benzene on water ice}
\author[Clark, Benoit, Van~de~Sande and Walsh]{Victoria H.J.\ Clark,$^{1}$\thanks{Contact email: \href{mailto:v.clark.17@ucl.ac.uk}{v.clark.17@ucl.ac.uk}}
David M.\ Benoit,$^{2}$\thanks{Contact email: \href{mailto:d.benoit@hull.ac.uk}{d.benoit@hull.ac.uk}}
Marie Van de Sande,$^{3,4}$\thanks{Contact email: \href{mailto:mvdsande@strw.leidenuniv.nl}{mvdsande@strw.leidenuniv.nl}}
and Catherine Walsh$^{3}$\thanks{Contact email: \href{mailto:c.walsh1@leeds.ac.uk}{c.walsh1@leeds.ac.uk}}\\
$^{1}$ Department of Physics and Astronomy, University College London, London, WC1E 6BT, UK \\
$^{2}$ E.A.\ Milne Centre for Astrophysics, University of Hull, Hull, HU6 7RX, UK\\
$^{3}$School of Physics and Astronomy, University of Leeds, Leeds, LS2 9JT, UK \\
$^{4}$Leiden Observatory, Leiden University, P.O. Box 9513, 2300 RA Leiden, The Netherlands}

\date{Last updated xx yy zz; in original form xx yy zz}

\pubyear{xxx}

\begin{document}
\label{firstpage}
\pagerange{\pageref{firstpage}--\pageref{lastpage}}
\maketitle

\begin{abstract}
In this paper we provide a highly accurate value for the binding energy of benzene to proton-ordered crystalline water ice (XIh), as a model for interstellar ices. We compare our computed value to the latest experimental data available from temperature programmed desorption (TPD) experiments and find that our binding energy value agrees well with data obtained from binding to either crystalline or amorphous ice. Importantly, our new value is lower than that used in most astrochemical networks by about nearly half its value. We explore the impact of this revised binding energy value for both an AGB outflow and a protoplanetary disk. 
We find that the lower value of the binding energy predicted here compared with values used in the literature (4050 K versus 7587 K) leads to less depletion of gas-phase benzene in an AGB outflow, and leads to a shift outwards in the benzene snowline in the midplane of a protoplanetary disk. Using this new value, the AGB model predicts lower abundances of benzene in the solid phase throughout the outflow. The disk model also predicts a larger reservoir of gas-phase benzene in the inner disk, which is consistent with the {recent detections of benzene for the first time in protoplanetary disks with JWST}.

\end{abstract}

\begin{keywords}
ISM:molecules -- ISM: dust -- molecular data -- astrochemistry
\end{keywords}

\section{Introduction}
{\citet{cernicharo_2001} first detected benzene in a planetary nebula (CRL618) at the start of this century. Although difficult to observe, benzene has also since been detected in a number of other astronomical environments such as post-AGB objects in the small Magellanic cloud (SMC) \citep{Kraemer:2006}, circumstellar envelopes of carbon-rich evolved stars \citep{Malek:2012} and the comae of comets and asteroids \citep{schuhmann_2019}. Closer to Earth, benzene has also been detected in the atmosphere of Saturn \citep{Koskinen_2016}, in Titan's atmosphere \citep{waite_2007} and has also long been identified as a component of meteoritic chondrites \citep{delsemme_1975}. 

Thanks to high-sensitivity observations with JWST (James Webb Space Telescope), {benzene has now also been detected for the first time in the inner regions of two protoplanetary disks around low-mass stars, with a high column density of 28\% and 68\% that of \ce{C2H2}, respectively  \citep{Tabone2023,Arabhavi_2024}}. 
{These new detections of abundant benzene in protoplanetary disks have} prompted renewed interest in this molecule, well motivating the need for the quantification of fundamental data pertinent to benzene chemistry in such environments, such as its binding energy to astrophysical ices and corresponding spectral features in the solid phase. {Indeed, a number of recent studies \citep{Ferrero_2020, Tinacci_2022, Bovolenta_2022} and reviews \citep{Zamirri_2019} have highlighted the importance of accurate binding energies for astrochemical simulations of smaller adsorbed molecules.} Furthermore, a suitable radio proxy for benzene has been identified as cyanobenzene \citep{Cooke_2020}, due to its rapid reaction with CN even at low cold ISM temperatures. Indeed cyanobenzene was first detected using in TMC-1 by \citet{McGuire_2018} and has now also been further confirmed in pre-stellar sources \citep{Burkhardt_2021}.

Benzene is also becoming a more popular component of kinetic reaction networks in dust grain models, since \citet{jones_2011} showed it could be formed through a barrierless reaction. However, an accurate binding energy data is key to providing a realistic account of the residence time of benzene on ice and describing the desorption process. Currently, the most popular value of benzene--ice binding energy used in those models (e.g., Rate12, \citeauthor{McElroy2013}~\citeyear{McElroy2013}) originates from an additive estimation by \citet{Garrod:2006}. In the present study, we suggest a new value obtained through high-level ab initio modeling of the binding of benzene to an ordered ice surface. We also compare our data to the latest temperature programmed desorption (TPD) experimental values \citep{thrower_2009, Stubbing:2019}.

We recently showed \citep{Clark_2019} the feasibility of a molecular arrangement where benzene acts as a hydrogen-bond acceptor to the dangling O--H bonds of the ice surface. Indeed, such $\pi$-hydrogen bonding was suggested already by \citet{silva_1994} and has also been identified for benzene--water clusters \citep{engdahl_1985, gotch_1992, suzuki_1992, benoit_1998}. We also explored the influence of such a binding mode on the vibrational spectrum of benzene \citep{Clark_2021} and showed how our ordered ice surface model was able to rationalise the laboratory-based infrared observations. The present study builds on these findings and has two aims: firstly, we use our surface model to provide a new reliable upper bound (i.e.\ largest) value for the binding energy of benzene on a water ice surface. Secondly, we explore the effects of using this revised binding energy for two astrophysically-relevant scenarios, namely AGB outflows and protoplanetary disk models. 

Our study is organised as follows: in Sec.~\ref{sec:binding_e}, we describe a hybrid approach to compute the binding energy of benzene adsorbed on a ferroelectric proton-ordered hexagonal crystalline water ice (XIh) surface and compare our values to the latest available experimental data. This ferroelectric surface provides us with a model of the most favourable (i.e.\ highest possible) binding energy for an adsorbed molecule on ice. Our computational details are outlined in Sec.~\ref{sec:compdet}. In Sec.~\ref{sec:results}, we present our new binding energy value for this system (along with derived data) and computational results exploring the implications of a weaker binding energy for astrophysical situations. Our conclusions can then be found in Sec.~\ref{sec:conc}.

\section{Surface binding energy model}
\label{sec:binding_e}
We use an ordered model for crystalline interstellar ice, built from the basal plane surface of ferroelectric water ice XIh, as described in our earlier work \citep{Clark_2021}. 
The binding energies for the benzene to the ice surface are computed using a technique that mixes periodic density functional theory (DFT) with coupled-cluster singles and doubles with perturbative triples (CCSD(T)) calculations. The scheme, referred later on as DFT/CCSD(T), is a version of the ``our own n-layered integrated molecular orbital and molecular mechanics" (ONIOM) approach developed by Morokuma and collaborators (see \citealt{dapprich_1999}). We use the optimised geometry of benzene molecule adsorbed on the \emph{topmost} surface layer of the ice XIh obtained in \cite{Clark_2021} (details in Sec.~\ref{sec:compdet}). The system is then divided into multiple parts as shown in Figure \ref{fig:interactionsplit}.

\begin{figure}
\resizebox{\columnwidth}{!}{
 \includegraphics{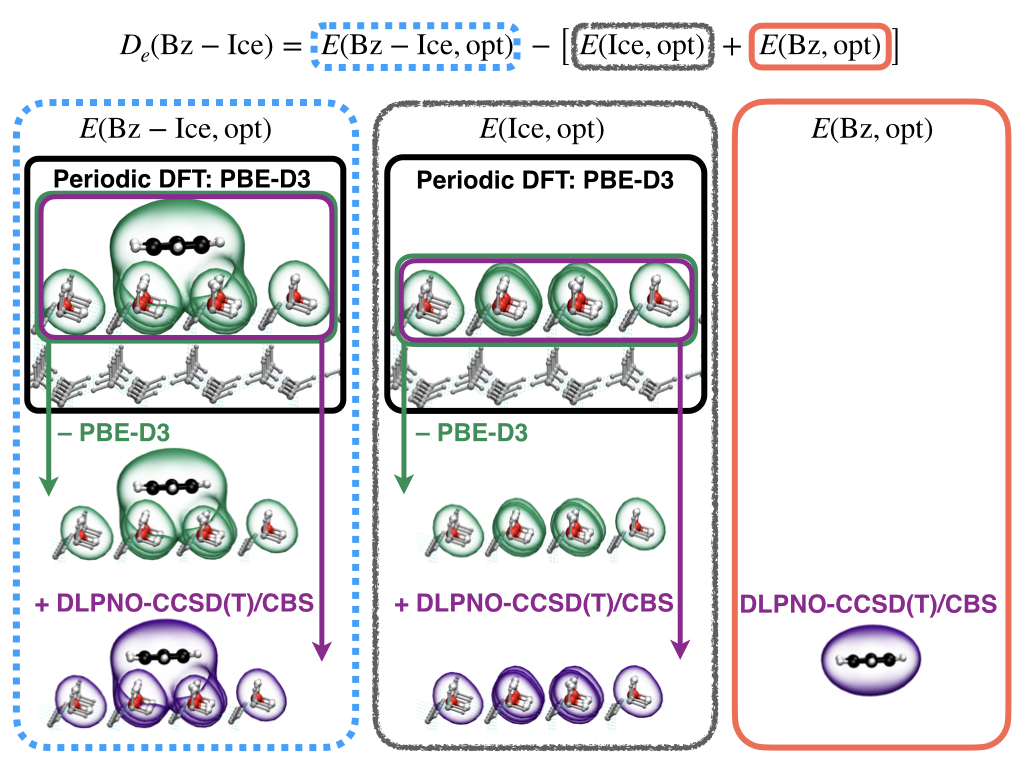}
}
\caption{Representation of ONIOM approach used to compute the adsorption energy of benzene on an ice surface model. Each component of the energy expression at the top is decomposed into up to three calculations: two periodic DFT calculations (full system and model) and one DLPNO-CCSD(T) calculation (model). The full system optimised at the PBE-D3/MOLOPT-TZV2P is shown at the top of each panel. Note that the benzene component is simply computed directly at the DLPNO-CCSD(T)/CBS level using an optimised PBE-D3/MOLOPT-TZV2P geometry. PBE-D3 calculations are indicated using a green colour and the DLPNO-CCSD(T) calculations using a purple colour.}
\label{fig:interactionsplit}
\end{figure}

Following the framework of the ONIOM approach, each component of the binding energy is systematically improved using an efficient implementation of CCSD(T) theory, the DLPNO-CCSD(T)/CBS approach (see Sec. \ref{sec:compdet} for details).

We define the binding energy of a benzene molecule as a negative quantity,
\begin{eqnarray}
D_e{_{(\mathrm{Bz-Ice})}}=E_{(\mathrm{Bz-Ice,opt})}-\left[E_{(\mathrm{Ice, opt})}+E_{(\mathrm{Bz,opt})}\right]\label{eq:bindingE},
\end{eqnarray}
where the $\mathrm{opt}$ label refers to an optimisation at the PBE-D3/MOLOPT-TZV2P level of theory, $E_{(\mathrm{Bz-Ice,opt})}$ refers to the bound benzene on ice, $E_{(\mathrm{Ice, opt})}$ refers to the pure ice surface and $E_{(\mathrm{Bz,opt})}$ refer to the gaseous benzene structure. Each component on the right of Equation \eqref{eq:bindingE} above is then computed using a two-level ONIOM approach such that,
\begin{eqnarray}
E_{(\mathrm{Bz-Ice,opt})}&=&E_{(\mathrm{Bz-Ice,opt})} ^\mathrm{low}-\left[E_{(\mathrm{model})} ^\mathrm{low}+E_{(\mathrm{model})} ^\mathrm{high}\right]\label{eq:bzice},\\
E_{(\mathrm{Ice, opt})}&=&E_{(\mathrm{Ice,opt})} ^\mathrm{low}- \left[ E_{(\mathrm{ice~model})} ^\mathrm{low}+E_{(\mathrm{ice~model})} ^\mathrm{high} \right] \label{eq:iceonly},\\
E_{(\mathrm{Bz,opt})}&=&E_{(\mathrm{Bz,opt}),} ^\mathrm{high}\label{eq:bzonly}
\end{eqnarray}
where here, the level of theory denoted ``$\mathrm{low}$" is PBE-D3/MOLOPT-TZV2P and the level of theory denoted ``$\mathrm{high}$" is DLPNO-CCSD(T)/CBS. Note that to compute the latter, we use the EC extrapolation approach of \citet{jurecka_2006} along with the aug-ano-pVnZ basis sets (see more details in section \ref{sec:compdet}). The overall benzene--ice system is split into a region of interest (the model system) and the rest of the system (see also Fig. \ref{fig:interactionsplit} for a visual representation of the ONIOM decomposition used). The model system in Equation \eqref{eq:bzice} is defined as the benzene molecule and all complete water molecules within 5~\AA\ of its centre of mass. Those same ice molecules also make up the ``ice model" system in Equation \eqref{eq:iceonly}, but this time with the relaxed geometry they adopt in a fully optimised ice surface. Finally the last component of the interaction energy (Equation \eqref{eq:bzonly}) is simply obtained from a DLPNO-CCSD(T)/CBS calculation for the fully optimised geometry of benzene at the PBE-D3/MOLOPT-TZV2P level of theory.

We also include zero-point energy (ZPE) corrections using the harmonic frequencies computed at the PBE-D3/MOLOPT-TZV2P ($\mathrm{low}$ level) for all three components of Equation \eqref{eq:bindingE}. Each of these corrections are then scaled using a factor of $1.0160$ recommended by \citet{Kesharwani_2014} to account for anharmonicities in ZPE calculations (note: that value was originally derived for PBE/def2-TZVPP calculations).

\section{Computational details}
\label{sec:compdet}
\subsection{Ice XIh model}
\label{sec:icedet}
As in our previous study (see \citet{Clark_2021} for more details), we use an ice surface obtained at the PW91/PW(350~Ry cutoff) level of theory by \citet{hirsch_2004}, which is repeated to create a $6\times 3 \times 2$ supercell. Our final model contains 36 unit cells of XIh ice in the slab, 288 H$_2$O molecules, with 4 double-layers of water molecules. The surface is chosen along the $c$ direction with an increased  lattice constant ($c=34.6716$~\AA) to accommodate the adsorption of a benzene molecule. The surface is relaxed such that the topmost water double-layer (72 H$_2$O molecules) is optimised, while the coordinates of the bottom three double-layers (216 H$_2$O molecules) of ice remains fixed. We use surface periodic boundary conditions in the $x, y$ ($a, b$) directions, while the $z$ ($c$) direction is treated non-periodically.

\subsection{DFT calculations}
All PBE-D3 calculations were performed using the Gaussian plane waves (GPW) method implemented in the QUICKSTEP module \citep{goedecker_1996, vandevondele_2005, perdew_1996} 
of CP2K (v4.1 and v6.1) \citep{vandevondele_2007, naumkin_1995, chergui_1996, goedecker_1996, lippert_1997}. The valence electrons are described using a TZV2P-MOLOPT-GTH basis set \citep{dunning_1971, vandevondele_2007, krack_2005}, along with an auxiliary plane-wave cutoff of 300~Ry. { The MOLOPT basis set family is optimised to reduce basis-set superposition error (BSSE) in periodic calculations (see \citet{vandevondele_2007}), as this can often be an issue for binding energy calculations, and has been shown to be roughly equivalent to a standard quadruple-zeta basis in terms of BSSE error.} The core electrons are represented using Goedecker-Teter-Hutter (GTH) pseudo potentials \citep{krack_2005}.

$XY$ periodicity was used for all calculations combined with an analytical Poisson solver for the electrostatic energy. The wave function convergence was set to 1.E-7~a.u.\footnote{Note that here ``a.u."~refers to atomic units while later on ``au" means astronomical units.} for all calculations. We use identical geometry convergence criteria to \citet{Clark_2021}. The exchange-correlation functional is that derived by Perdew, Burke and Ernzerhof (PBE) \citep{perdew_1996} and we account for dispersion interactions using the DFT-D3 correction scheme of \citet{Grimme_2010} based on a damped atom-pairwise potential and three-body $C_9$ corrections (ATM correction, \citealt{Moellmann_2014}). {Note that the DFT-D3 methodology was parameterised for basis sets of quadruple-zeta quality, assuming that those are near the basis-set limit \citep{Grimme_2010}.} The geometry of the adsorbed benzene on ice, along with selected structural parameters have been reported in \citet{Clark_2019} and a coordinate file was provided in the SI of \citet{Clark_2021}. 

\subsection{DLPNO-CCSD(T)/EC2-CBS calculations}
\label{sec:dlpnodet}
The complete basis set extrapolation of the domain-based local-pair natural orbital coupled-cluster singles doubles and perturbative triples (DLPNO-CCSD(T)) energy was performed using the ORCA 3.0.3 suite of programs. Note that this particular version of ORCA, the (T) implementation uses a ``semi-canonical'' approximation to compute the perturbative triples correction, also known as T0 correction. The expression used for the EC2-CBS extrapolation, due to \citet{jurecka_2006} is
\begin{eqnarray}
E(\textrm{DLPNO-CCSD(T)/EC2-CBS(X,Y)}) &\approx& E(\mathrm{SCF;Y}) + E(\mathrm{DLPNO-CCSD(T);X})\nonumber\\
{}&& + E(\mathrm{MP2;}\infty) - E(\mathrm{MP2;X}),
\end{eqnarray}
where $\mathrm{X}$ and $\mathrm{Y}$ are the cardinal number of each basis set used. For our binding energy calculations, we used the aug-ano-pVDZ basis set together with the aug-ano-pVTZ basis set, and thus $\mathrm{X}=2$ and $\mathrm{Y}=3$ respectively. We also compared to a higher level aug-ano-pVTZ and aug-ano-pVQZ basis set combination ($\mathrm{X}=3$ and $\mathrm{Y}=4$) and a near-complete basis approach with the aug-ano-pVQZ and aug-ano-pV5Z basis set combination ($\mathrm{X}=4$ and $\mathrm{Y}=5$). %

The MP2 energy is extrapolated using
\begin{eqnarray}
E(\mathrm{MP2;}\infty)=\frac{\mathrm{X}^\beta\cdot E(\mathrm{MP2,X})-\mathrm{Y}^\beta\cdot E(\mathrm{MP2,Y})}{\mathrm{X}^\beta -\mathrm{Y}^\beta},
\end{eqnarray}
where $\beta=2.41$, value optimised by \citet{neese_2009, neese_2011}.

In order to accelerate the calculations, we used both the RI-JK approximation \citep{weigend_2008} (with a cc-pVQZ/JK auxiliary basis set, \citealt{dunning_1989}) and the RI-MP2 approach \citep{weigend_1997} (with a aug-cc-pVQZ/C auxiliary basis set, \citealt{kendall_1992}). All calculations used the verytightscf convergence criterion of ORCA \citep{neese_2012}. All DLPNO-CCSD(T) calculations used the default ORCA criteria for the PNO generation (NormalPNO) unless stated.

 \section{Results and Discussion}
 \label{sec:results}
 \subsection{Binding energy} 
\label{sec:binding_val}

The highly dipolar surface arrangement chosen for the ice surface provides a best-case scenario model for the binding of benzene to an ice surface since it maximises the number of possible hydrogen-bond donors at the surface. It is expected that a realistic non proton-ordered ice surface (Ih or ASW, for example) would have fewer binding sites available for benzene, although an interesting striped proton ordering has been suggested for ice Ih surface \citep{buch_2008} and dangling OH bonds are a well-characterised feature of ASW \citep{mccoustra_1996}. Therefore the results obtained for our model can be considered to provide an upper-bound (i.e.\ most binding) case for the binding energy of benzene on a water ice surface, which is the aim of our study. 

\subsubsection{Cluster benchmarks}
\label{sec:clustbench}
\emph{Benzene-water}\\
In order to assess the overall accuracy of the hybrid approach proposed in section \ref{sec:binding_e}, we first perform calculations on the benchmark benzene--H$_2$O system using the geometry reported in the S22 database initially developed by \citet{jurecka_2006}. Our results are summarised in the top panel of Table \ref{tab:bzbind}. 

\begin{table}
 \centering
 \caption{Computed and measured binding energies for benzene-H$_2$O and the water dimer. Values are given in kJ/mol with uncertainties if available. Brackets indicate ``tightPNO" values. All calculations use the cluster geometries from the S22 database (obtained from www.begdb.com \citep{rezac_2008}).
}
 \begin{tabular}{@{} llr @{}} %
 \hline
 \hline
System&Method & Binding energy [kJ/mol]\\
\hline
Benzene--H$_2$O cluster&&\\
&PBE-D3/MOLOPT-TZV2P&$-15.08$\\
&DLPNO-CCSD(T)/EC2-CBS(2,3)&$-13.52~(-13.44)$\\
&DLPNO-CCSD(T)/EC2-CBS(3,4)&$-13.29~(-12.85)$\\
&DLPNO-CCSD(T)/EC2-CBS(4,5)&$-13.66~(-13.08)$\\
&CCSD(T)/CBS$^a$&$-13.77$\\
\\
\cline{2-3}\\
&\textit{Experimental data}&\\
&$D_0$ \citep{gotch_1992}&$-6.82 \leq D_0 \leq -11.63$\\
&$D_0$ \citep{cheng_1995}&$-9.4\pm 1.2$\\
&$D_0$ \citep{courty_1998}&$-10.2\pm 0.4 $\\
&Estimated$^b$ $D_e$&$-13.1\pm 0.4 $\\
\hline
H$_2$O--H$_2$O cluster&&\\
&PBE-D3/MOLOPT-TZV2P&$-23.33$\\
&DLPNO-CCSD(T)/EC2-CBS(2,3)&$-20.35~(-20.57)$\\
&DLPNO-CCSD(T)/EC2-CBS(3,4)&$-20.17~(-20.58)$\\
&DLPNO-CCSD(T)/EC2-CBS(4,5)&$-20.36~(-20.63)$\\
&CCSD(T)/CBS$^a$&$-21.21$\\
\\
\cline{2-3}\\
&\textit{Experimental data}&\\
&$D_0$ \citep{rocher-casterline_2011}&$-13.20\pm 0.05$\\
&Estimated$^c$ $D_e$&$-20.80\pm 0.06$\\
\hline
\hline
 \end{tabular}
{\flushleft \footnotesize
$^a$ Revised value from \citet{takatani_2010}; \\
$^b$ Value estimated from the experimental $D_0$ measurements for benzene--H$_2$O and benzene--D$_2$O from \citet{courty_1998} and the zero-point energy values computed for the same systems using rigid-body diffusion Monte Carlo calculations by \citet{benoit_2000}.\\
$^c$ Value estimated from the experimental $D_0$ measurements for the dissociation energy of the water dimer from \citet{rocher-casterline_2011} and the zero-point energy values computed by \citet{shank_2009}\\
}
 \label{tab:bzbind}
\end{table}

We see that, compared to the revised CCSD(T)/CBS reference energy value of \citet{takatani_2010} for benzene--H$_2$O, PBE-D3 over-estimates the binding energy by about 10\%. The slight over-estimation of binding energy by the D3 correction has already been observed \citep{reckien_2014} for benzene adsorption on metal surfaces, for example. Moreover, our chosen basis set is also known to cause some overbinding for ice Ih (see \citet{brandenburg_2019a}, footnote 151). However, given the low computational cost of this approach, this result is still very reasonable. Next, we see that the DLPNO-CCSD(T)/EC2-CBS extrapolation approach in its various format (2,3), (3,4) or (4,5) all lead to good agreement with the extrapolated results of \citet{takatani_2010}. The ORCA team recommends the usage of a ``TightPNO" \citep{laikos_2015} criterion when computing interaction energies for weakly bound systems (those values are reported in brackets in our Table \ref{tab:bzbind}). We see here that the (2,3) extrapolation is in surprisingly good agreement with the full CCSD(T)/EC2-CBS data, with a deviation of only 2\%, regardless of the PNO criterion. Keeping in mind that our approach relies on an MP2 estimation for the correlation extrapolation, it is very likely that this outcome is the result of fortuitous error cancellation for this combination of basis sets and/or extrapolation parameters. The (3,4) extrapolation performs slightly worse than (2,3), but is still only 3\% (7\%) away from the reference and the (4,5) extrapolation is a mere 0.7\% (5\%) above the reference result. 

Another comparison point is provided by the work of \citet{brandenburg_2019} 
who used quantum diffusion Monte-Carlo to determine the binding energy of a water molecule to benzene. They compute a binding energy of $-13.1\pm 0.5$ kJ/mol for their ``2-leg" configuration, in very good agreement with the values reported in Table \ref{tab:bzbind}. 

One interesting conclusion from these results is that, for this particular system, the usage of TightPNO increases computational cost with limited accuracy increase and thus the standard criteria appear to provide a better cost/performance balance in our case. Furthermore the computational cost of the (2,3) EC extrapolation are much lower than those of the (4,5) extrapolation but still lead (fortuitously) to results of similar quality. { Finally, we also see that a good agreement with the reference BSSE-corrected data can be achieved without any BSSE corrections (see also supplementary information).} While this is not likely to be a conclusion of wide-ranging application, since the larger extrapolations and tighter selection criteria are usually more reliable, it does provide a practical solution for the estimation of binding energies beyond the DFT-D3 level for large systems.

A further measure of the accuracy of the computed binding energies is to compare those to the available experimental data, rather than other high-level calculations. The measurement of the binding energy of the benzene--water complex has been performed using a variety of techniques over the years, leading to values ranging from $-6.8$ to $-11.6$ kJ/mol. A selection of these values are reported in the middle of Table \ref{tab:bzbind} under the section ``\emph{experimental data}". The measurement by \citet{courty_1998} 
is the most accurate to date and also agrees with the error bars and estimations of previous studies. However, the measured experimental values ($D_0$) cannot be directly compared with the theoretical binding energies ($D_e$), as the former include the zero-point vibrational energy of the complex. In their study, \citet{courty_1998} used their $D_0$ measurements along with early rigid-body diffusion Monte Carlo (RB-DMC) calculations by \citet{gregory_1996} to estimate $-12.89 \leq D_e \leq 14.80$~kJ/mol. To provide a slightly tighter $D_e$ estimate, we combined the experimental $D_0$ measurements for benzene--H$_2$O and benzene--D$_2$O from \cite{courty_1998} and the zero-point energy (ZPE) values computed for the same systems using improved RB-DMC calculations by \citet{benoit_2000} on an accurate model potential. This leads to a value of $-13.1\pm 0.4$ kJ/mol, in very good agreement with the reference CCSD(T)/CBS data of \citet{takatani_2010}, the DMC estimation of \citet{brandenburg_2019} and our DLPNO-CCSD(T)/EC2-CBS values, given the nature of the ZPE calculations. This suggests that our value ($-13.1\pm 0.4$ kJ/mol) is an accurate experimental $D_e$ value for the benzene water cluster.
\\

\emph{Water dimer}\\
In order to explore if our previous conclusions extends to water (and ice), we briefly investigate the binding energy of the water dimer (also from the S22A benchmark set). Our results are summarised in the lower section of Table \ref{tab:bzbind}. We observe here again that PBE-D3 over-estimates the binding by a similar amount to the benzene--water case (10\%). The extrapolated DLPNO-CCSD(T)/EC2-CBS data is also interesting as they further demonstrate a surprisingly good agreement of the (2,3) extrapolation, within 4\% (3\%) deviation from the reference value. As was the case earlier, the (3,4) data is slightly worse than (2,3) with a deviation of 5\% (3\%) at a much increased computational time. Finally, the most accurate (4,5) extrapolation is of similar quality to the previous results obtained with (2,3) and lead to a deviation of 4\% (3\%) from the reference value. 
We note that the change due to the usage of the ``TightPNO" criterion in the DLPNO-CCSD(T) calculation is marginal, although it does seem to compensate for the level of extrapolation with all three scheme leading to a corrected value of $-20.6$~kJ/mol. 

An appropriate comparison with experimental data also requires the subtraction of the ZPE from the excellent experimental measurements of $D_0$ from \citet{rocher-casterline_2011}. Here, we use the ZPE estimate computed using diffusion Monte-Carlo by \citet{shank_2009} on a highly-accurate model potential. We observe that there is good agreement between the estimated ``experimental" $D_e$ and both the reference CCSD(T)/CBS value and our DLPNO-CCSD(T)/EC2-CBS estimations{, without any BSSE correction (see also supplementary information)}.

These benchmark tests provide a suitable validation of our approach and ascertain the accuracy of both PBE-D3 (about 10\% overestimation) and the DLPNO-CCSD(T)/CBS(2,3) approach (about 4\% deviation) as a way of systematically computing the binding energy of molecules interacting with water. 

\subsubsection{Benzene on ferroelectric ice XIh model}
Our results are shown in Table \ref{tab:bzicebind}. We observe, here too, that PBE-D3 seems to over-estimate the binding of benzene on ice, if we compare to the values obtained by \citet{sharma_2016} for the their M06-2X DFT model, for example. This is to be expected, based on both cluster energy comparisons described earlier (see section \ref{sec:clustbench}) and existing literature (\citet{reckien_2014}, for example). Our benchmark cluster data reported in section \ref{sec:clustbench} suggests that an error on the PBE-D3 values of 10\% is appropriate, which accounts for some of the deviation from the values computed by \citet{sharma_2016}.

\begin{table}
 \centering
 \caption{Computed and measured binding energies for benzene adsorbed on a ferroelectric proton-ordered hexagonal crystalline water ice (XIh) surface. Values are given in kJ/mol with uncertainties if available. $^a$ \citet{sharma_2016} reported for the geometry closest to our XIh surface model and their largest quantum subsystem (structure A1 in their study). $^b$ Estimated zero-point energy correction based on scaled harmonic frequency calculations, details given in the text.}
 \begin{tabular}{@{} llr @{}} %
 \hline
 \hline
System&Method & Binding energy [kJ/mol]\\
\hline
Benzene--ice (XIh)&&\\
&PBE-D3/MOLOPT-TZV2P&$-49\pm 5$\\
&DLPNO-CCSD(T)-hybrid/CBS(2,3)&$-36 \pm 1$\\
&DLPNO-CCSD/cc-pVTZ:FF$^a$& $-43.4$\\
&M06-2X/6-31++G(d,p):FF& $-43.4$\\
&M06-2X/6-31++G(d,p)$^a$&$-41.5$\\
\\
Suggested $D_0$&DLPNO-CCSD(T)-hybrid/CBS(2,3)+ZPE$^b$&$-34 \pm 1$\\
\\
&\textit{Experimental data}&\\
&UMIST database Rate12&$-63.2$\\
&\citet{thrower_2009} \ (TPD@ASW)&$-41.0\pm0.5$\\
&\citet{Stubbing:2019} \ (TPD@ASW)&$-42\pm6$\\
&\citet{Stubbing:2019} \ (TPD@CI)&$-39\pm3$\\
\hline
\hline
 \end{tabular}
 \label{tab:bzicebind}
\end{table}

To improve on our PBE-D3 value, while conserving the collective effects of the ice surface model, we use the method described in section \ref{sec:binding_e} and obtain values that are in good agreement with those of \citet{sharma_2016} for a similar model (also using a hybrid model, but not including periodicity effects explicitly). Our benchmark cluster data reported in section \ref{sec:clustbench} indicates that an error of 4\% is appropriate for our approach, leading to a binding energy estimate of $D_e=-36 \pm 1$~kJ/mol. 

Our value is slightly higher (i.e.\ less negative) than the results reported by \citet{sharma_2016} both for their DFT-based approaches (M06-2X with or without hybrid corrections) and their CCSD-based hybrid estimate. To explain some of this discrepancy, we note that \citet{mackie_2010} have shown that M06-2X can over-estimate the binding energy for the benzene-water system, despite M06-2X being overall an accurate functional for weak interactions \citep{mardirossian_2017}. \citet{sharma_2016} show a good agreement between M06-2X and DLPNO-CCSD binding energies for their models, but a detailed study by \citet{crittenden_2009} showed that neglecting (T) correction usually lead to underbinding at the basis-set limit. For triple-zeta basis sets, without any basis correction, even CCSD(T) is shown to over-estimate the binding energy of benzene water \citep{feller_1999}. To gain further insights into the accuracy of the two approaches, we used M06-2X/6-31++G(d,p) and DLPNO-CCSD/cc-pVTZ to compute the binding energy of the benzene-water benchmark system described in Sec.\ref{sec:clustbench}. Our results showed that both method likely overestimate the binding energy (by 18\% for M06-2X, with $-16.9$~kJ/mol, and by 10\% for DLPNO-CCSD, with $-15.1$~kJ/mol). %
{ Finally, we also investigated the impact of BSSE on the computed binding energy for benzene on ice XIh and found it to be negligible within the uncertainty of our approach ($\ll$~1~kJ/mol, see supplementary information).}
Thus we conclude that our binding energy estimate is more representative of the benzene binding energy to crystalline water ice (XIh).

In order to be able to compare with experimental data, we also need to include zero-point energy (ZPE) correction for this system. This is somewhat more problematic as a full optimisation of the benzene+ice system is not an easy task. Here we explore two ways of doing this. First, the simplest approach is to only consider the translation and rotation (TR) modes of benzene on ice and use those to compute the ZPE. Those modes are traditionally removed through projection in most codes, but can easily be extracted from the unpurified Hessian matrix. The difference between the harmonic TR modes for benzene ice and benzene gives us a ZPE of 154~\icm (1.85~kJ/mol). In order to account for anharmonicty, we use here the scaling approach of \citet{Kesharwani_2014} and use a scaling factor of 1.016 for PBE, leading to a ``corrected" ZPE of 157~\icm (1.88~kJ/mol). An alternative approach is to consider both the benzene and the first ice layer, directly in contact with the adsorbate. This option is computationally more expensive and requires careful control over the optimisation of the equilibrium geometry for both benzene on ice and for the isolated ice model. After frequency projection (in this case, the overall translations and rotations modes are not relevant for intermolecular ZPE), we obtain a ZPE of 161~\icm (1.93~kJ/mol), in very close agreement with our earlier simpler approach. The anharmonic value obtained using the PBE scaling factor of 1.016 defined above leads to a ``corrected" ZPE of 164~\icm (1.96~kJ/mol).

{ The work from \citet{Slipchenko_2008} shows a similar trend in their ZPE corrections, where their benzene--water binding energy is only increased by about 1~kcal/mol (4.2~kJ/mol) from their computed $D_e$ value. \citet{feller_1999} also suggests a similar value of 1~kcal/mol (4.2~kJ/mol) value for the ZPE correction of the same complex. If we account for anharmonic corrections and keep in mind that the values reported in both studies are for a \emph{single} water molecule binding to a benzene molecule, those estimations broadly agree with ours. Indeed, the lower mobility of the water molecules in the ice lattice is likely causing the low impact of benzene on the benzene--ice ZPE correction.}

Our estimate for the experimental binding energy is thus $D_0=-34 \pm 1$~kJ/mol, which is nearly half the value of the Rate12 UMIST database commonly used in astrochemical modeling (see also Table \ref{tab:bzicebind}). This new estimate is in good agreement with the TPD measurements of \citet{thrower_2009} ({$-41.0\pm0.5$~kJ/mol, recommended average value from their desorption study}) and Stubbing~\&~Brown \citep{Stubbing:2019} for benzene on ASW ($-42\pm6$~kJ/mol). We also note that the binding energy obtained from TPD of benzene on crystalline ice (CI) leads to a slightly lower value of $-39\pm 3$~kJ/mol \citep{Stubbing:2019}, also in close agreement with our estimate.
This implies either that the simple ordered model we developed gives a realistic description of the local environment experienced by benzene on ASW (i.e.\ on a local scale, the binding mode of benzene in ASW resembles that of ice XIh), or that the uncertainly in both experiments or our calculations is still too large to be able to differentiate between the two types of ices (crystalline XIh and amorphous ASW). We are currently exploring this avenue further and will report our findings in a future publication.

\subsection{Astrophysical implications}

To test the impact of the binding energy for benzene calculated here ($-34$~kJ/mol or a desorption energy of 4090~K), we adopt this value in a chemical model of a carbon-rich AGB (asymptotic giant branch) outflow \citep{VanDeSande2021} and the mid-plane of a protoplanetary disk around a T Tauri star \citep{Walsh2015}. We compare the results with the value listed in the Rate12 version of the UMIST Database for Astrochemistry \citep[UDfA, 7587~K;][]{McElroy2013}. 
Full details of both models are given in the provided references and we provide here some brief details only. 
Both models use the full UDfA gas-phase chemical network supplemented with gas-grain interactions (including accretion, thermal desorption, and non-thermal desorption driven by stellar and interstellar photons and cosmic-ray induced photons, and reactive desorption) and a sub-set of the grain-surface network (and associated gas-phase chemistry and gas-grain interactions) from \citet{Garrod2008}. 
The AGB outflow model also includes sputtering from grains as a desorption mechanism \citep[see][for further details]{VanDeSande2019}.

\subsubsection{Carbon-rich AGB outflow}

The AGB outflow model adopts parameters similar to the well-studied carbon-rich AGB star, IRC+10216: a mass-loss rate of $10^{-5}$ M$_\odot$ yr$^{-1}$, a constant outflow velocity of 15 km s$^{-1}$, a stellar temperature of 2000 K, and a temperature profile characterised by a power law with exponent 0.7 \citep{VanDeSande2021}.
We run four variations of the outflow model; one in which we assume that benzene is a daughter species only (with a zero initial abundance), and one in which benzene is assumed to be formed in the inner wind and is thus a parent species.  
The parents species and abundances are those of \citet{Agundez2010}.
The initial abundance of benzene when included as a parent species is $8.7 \times 10^{-7}$ with respect to \ce{H2}, or two orders of magnitude less than the abundance of the parent \ce{C2H2}, which is consistent with the model predictions of inner wind chemistry from \citet{Cherchneff2012}.  
Note that gas-phase benzene is yet to be detected in the outflow of an AGB star; hence, we use an optimistic value from the models for this case.
For each set of initial abundances, we also run a version of the model in which we include gas-phase and gas-grain chemistry only (as done in \citet{VanDeSande2019}), and one in which we also enable the processing of carbon-chain species to form refractory organics (as described in \citet{VanDeSande2021}).

Fig~\ref{fig:agb_results} shows the fractional abundance of benzene with respect to \ce{H2} in the gas-phase (solid lines) and ice phase (dashed lines) for the models in which benzene is a daughter species (top row) and in which benzene is a parent species (bottom row). 
The left-hand and right-hand plots show the model results excluding and including, respectively, the photo-processing of carbon-rich ices to form refractory organics.  
In the right-hand plots the fractional abundance of benzene `converted' to refractory form is shown by the dotted lines.
The results from the models that adopt the default UDfA value are shown in orange, and the model using the binding energy from this work are shown in blue. 

The models in which benzene is assumed to be a parent predict a much larger peak fractional abundance of gas-phase benzene in the outflow ($\sim 10^{-6}$ versus $\sim 10^{-8}$ with respect to \ce{H2}). 
The models show that if benzene is a parent species, it is able to persist in the gas-phase at (close to) the assumed initial abundance until a radius of $\approx 2 \times 10^{17}$~cm at which point it starts to be dissociated by interstellar UV photons. 
In the models in which benzene is a daughter species, the fraction abundance peaks at a radius of $\sim 10^{17}$~cm.  
The formation of gas-phase benzene in this case is via ion-molecule chemistry. 
\ce{C6H6} is produced by dissociative recombination with electrons by \ce{C6H7+}, which is formed by radiative association {of \ce{H2} with} \ce{C6H5+}}. 
\ce{C6H5+} formation is linked to the protonation of \ce{C2H2} to \ce{C2H3+}, which then reacts with \ce{C2H2} to form \ce{C4H3+} and subsequently \ce{C6H5+}.

In the models without photo-processing, the use of the lower binding energy has a negligible effect on the peak abundance and column density of gas-phase benzene both in the models in which it is a daughter species and in the models in which it is a parent species.
However, the peak fractional abundance of ice-phase benzene (i.e., benzene that is bound to the ice) drops by a factor of $\approx 2$ for the model in which benzene is a daughter, and by a factor of $\approx 4$ in the model in which benzene is a parent. 

In the models with photoprocessing, gas-phase benzene is similarly unaffected by the change in binding energy.  
The model adopting the lower binding energy predicts a lower peak fractional abundance of `refractory' benzene by a factor of $\approx 3$ for the case where benzene is a daughter, and by a factor of $\approx 4$ for the case where benzene is a parent. 
The use of the lower binding energy also delays the onset of freezeout (and thus conversion to `refractory' benzene) from a radius of $\approx 5\times 10^{15}$~cm to a radius of $\approx 2 - 3 \times 10^{16}$~cm. 
The lower binding energy for this model predicts significantly less benzene in the ice phase than that using the higher binding energy. 

To summarise, the impact of using a lower binding energy for benzene has negligible impact on the gas-phase benzene through the outflow, but lowers both the ice-phase and refractory-phase benzene.  
The model including the photo-processing of ices and the lower binding energy calculated here predicts negligible amounts of benzene in the ice phase.

\begin{figure}
    \centering
    \includegraphics[width=\textwidth]{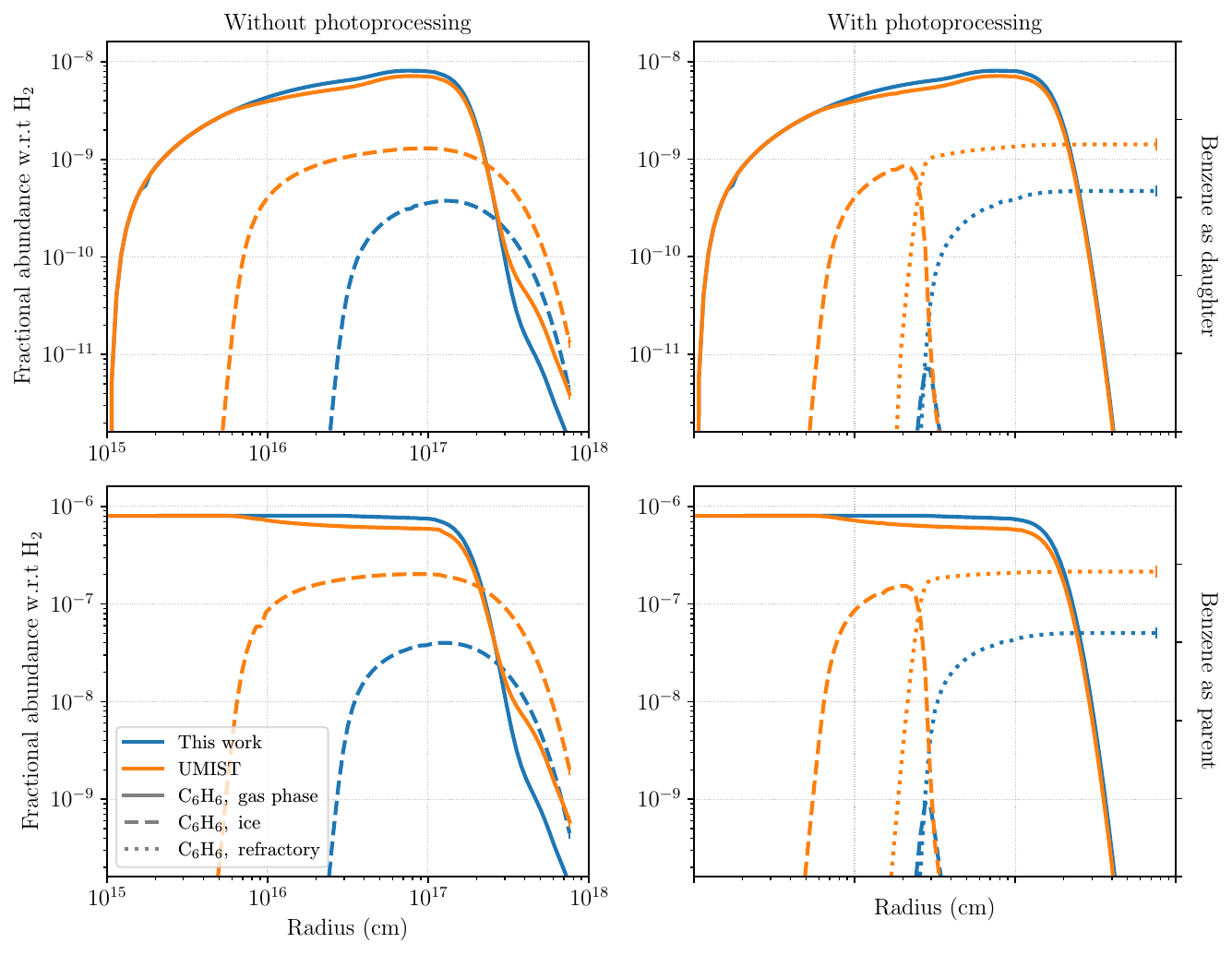}
    \caption{Fractional abundance (with respect to \ce{H2}) for gas-phase benzene (solid lines), ice phase benzene (dashed lines) and `refractory phase' benzene (dotted lines) for models in which benzene is assumed to be a daughter species (top row) and in which benzene is assumed to be a parent species (bottom row), and for models excluding (left) and including (right) the photo-processing of carbon-rich ices to refractory organics. The results when adopting the UDfA value of the benzene binding energy (7587~K )are given in orange, and those using the binding energy calculated in this work (4090~K) are given in blue. }
    \label{fig:agb_results}
\end{figure}

\subsubsection{Disk midplane of a T Tauri star}

The protoplanetary disk model is the T Tauri disk model described in \citet[][]{Walsh2015}.  
Because we are modelling the chemical structure of the warm and dense inner disk midplane which is well shielded from radiation, we include only gas-phase chemistry, and gas-grain interactions (accretion, and thermal and non-thermal desorption driven by cosmic-rays), that is, we do not include grain-surface reactions nor reactive desorption.  
The initial abundances adopted in this model are taken from a dark cloud chemical model with a temperature of 10~K, a number density of $10^{4}$~cm$^{-3}$, and a cosmic-ray ionisation rate of $1.3 \times 10^{-17}$~s$^{-1}$. 
Abundances are extracted at a time of $3 \times 10^{5}$~years. 
Note that the binding energy adopted for benzene in the dark cloud model is the default value from 
UdFA (7587~K). 
The initial abundance of benzene in this model is $2.6 \times 10^{-11}$ in the gas-phase and $6.5 \times 10^{-12}$ in the ice phase, with respect to total H nuclei density. 

\begin{figure}
    \centering
    \includegraphics[width=0.49\textwidth]{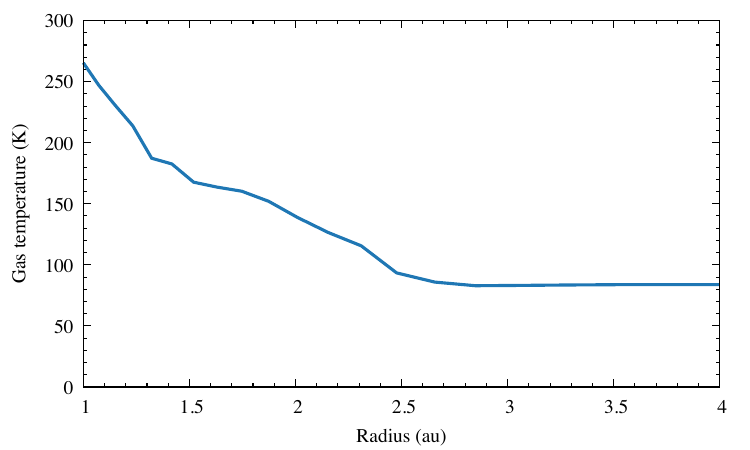}
    \includegraphics[width=0.49\textwidth]{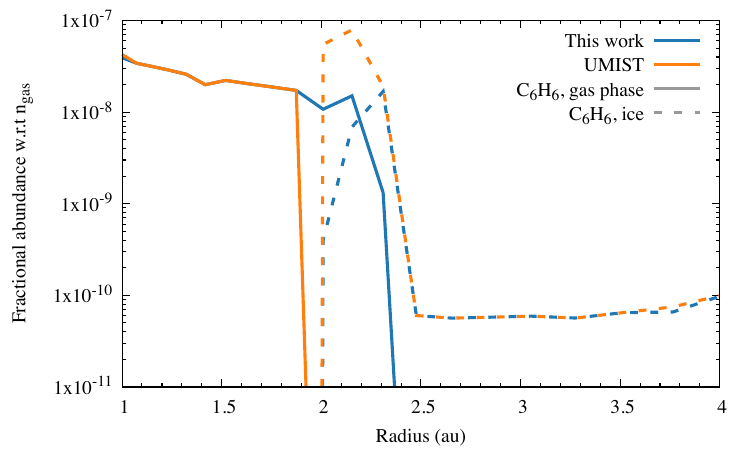}    
    \caption{Temperature profile (right) of the inner midplane for a disk around a T Tauri star (from the models presented in \citet{Walsh2015}). Fractional abundance (with respect to total gas density) for gas phase benzene (solid lines) and ice phase benzene (dashed lines) for a model in which the binding energy from this work was adopted (4090~K; blue) and for a model using the default UDfA value (7587~K; orange).}
    \label{fig:disk_results}
\end{figure}

The left-hand panel of Fig.~\ref{fig:disk_results} shows the temperature profile for the disk model from \citet{Walsh2015}.  
The temperature decreases as a function of radius from a value of 265~K at 1~au to a value of 85~K at 4~au.  
Note that this disk has two sources of heating in the inner midplane: passive heating from the central star and heating from accretion \citep[see][for more details on the physical model of the disk]{Nomura2007}.

The right-hand panel of Fig.~\ref{fig:disk_results} shows the fractional abundance of gas-phase and ice-phase benzene as function of radius.  
For the default UDfA value for the benzene binding energy, the benzene snowline lies at $\approx 1.8$~au.  
When adopting the value calculated here, the benzene snowline shifts outwards to $\approx 2.2$~au.  
Note that the apparent gap in the amount of benzene in the disk for the model adopting the UDfA value is an artefact of plotting the abundances on a log scale. 
The abundance of gas-phase benzene in this model drops quickly to negligible values; whereas the abundance of ice-phase benzene rises quickly from a negligible value in the  inner disk.
Hence, the model adopting the binding energy calculated in this work predicts a larger and more radially extended reservoir of gas-phase benzene in the inner disk.  
This is a similar trend to that found by \citet{woods_2007} who also explored the impact of benzene binding energy on the distribution and abundance of benzene in the inner disk, albeit for a different disk model and network to that used here.  
\citet{woods_2007} saw the benzene snowline move from $\approx 1$ to $\approx 2$~au when decreasing the binding energy of benzene from 7587~K to 4750~K.  
This lower value comes from estimates of the heats of adsorption of benzene to a graphite surface using gas-solid chromatography \citep{Arnett1988}, and is very close to the value calculated here.
Also evident in these results is the efficient formation of gas-phase benzene in the inner disk, increasing from an initial fractional abundance of $2.6 \times 10^{-11}$  to a value of $\approx 2 \times 10^{-8}$ with respect to total H nuclei density (note that in the inner disk, most of the hydrogen is in molecular form).  
This boost in the abundance of gas-phase benzene is due to efficient ion-molecule chemistry: \ce{C6H6} is formed via the recombination of \ce{C6H7+} with negatively charged grains which are the main charge carriers in this region of the disk, being some two to three orders of magnitude more abundant than electrons.  
\ce{C6H7+} formation, in turn, is dominated by the barrierless reaction between \ce{C3H4+} and \ce{CH3CCH} with the former produced predominantly by cosmic-ray-induced photoionisation of \ce{CH3CCH} and the latter from the recombination of \ce{C4H5+} with negatively charged grains. 
This mechanism of benzene formation in the inner disk is also described in \citet{woods_2007}. 

\section{Conclusion}
\label{sec:conc}

In this paper we have presented a reliable upper bound (i.e.\ largest) value for the binding energy of benzene on an ordered water ice surface ($-34$~kJ/mol or a desorption energy of 4090~K, when including zero-point energy corrections). Our estimation is based on a carefully-calibrated hybrid method that includes both a high-level treatment of electronic correlation, through the DLPNO-CCSD(T) approach, but also the influence of surface periodicity through Gaussian plane wave density functional theory. We have shown that our benzene-ice binding energy is different to what has been used in astrochemical models, sometimes by up to 50\%. We also compared our estimations with TPD desorption data. While two sets of experiments are performed on amorphous structured water (ASW) ice and one on crystaline water ice, we find that all three values agree well with our computed highly-ordered crystalline ice data. This implies that our simple ordered model describes realistically the local environment experienced by benzene on ASW (i.e.\ on a local scale, the binding mode of benzene in ASW resembles that of ice XIh). 

We also explored the influence of this revised binding energy on astrophysical models. Inputting our value into a model of the gas-grain chemistry occurring in an AGB outflow shows that using the lower value delays the onset of freezeout (and thus depletion) of benzene in the wind. When photoprocessing is included, the lower binding energy has the additional effect of limiting the abundance benzene in the ice phase to negligible values. 
The new binding energy also predicts a shift outwards in the position of the benzene snowline in the midplane of a protoplanetary disk by $\approx 0.4$~au (this work) to $\approx 1$~au \citep{woods_2007}. 
Hence, a lower binding energy predicts a larger reservoir of gas-phase benzene in the inner disk compared with the default UDfA value.  
{ We note here that gas-phase benzene has been detected in now two protoplanetary disks for the first time with JWST \citep{Tabone2023,Arabhavi_2024}}.  
The impact of this lower binding energy on the abundance and distribution of gas-phase benzene in the inner disk should be explored in future work, especially under carbon-rich conditions, given this new result from JWST. Based on our work, we recommend using this revised value ($-34$~kJ/mol or a desorption energy of 4090~K) for the binding energy of benzene on water ice in astrochemical models.

\section{Acknowledgements}
We thank Dr James Stubbing and Professor Wendy Brown (University of Sussex, UK) for generously providing experimental data used in this work. We also thank Dr Gerrit Brandenburg for helpful comments and suggestions regarding the binding energy computation. 
We acknowledge the Viper High Performance Computing facility of the University of Hull and its support team.
D.M.B.~acknowledges financial support from the Science and Technology Facilities Council (grant number ST/R000840/1). 
MVdS acknowledges support from the  European Union's Horizon 2020 research and innovation programme under the Marie Skłodowska-Curie grant agreement No 882991 and the Oort Fellowship at Leiden Observatory. 
C.W.~acknowledges financial support from the University of Leeds, the Science and Technology Facilities Council, and UK Research and Innovation (grant numbers ST/X001016/1 and MR/T040726/1).

\section{Data Availability}
All data used for this study is provided in the paper or references herein.

\bibliographystyle{mnras.bst}
\bibliography{paper}

\label{lastpage}

\end{document}